\documentclass[12pt]{article}
\usepackage{graphicx}
\usepackage{amssymb}
\usepackage{amsmath}
\usepackage{color}
\numberwithin{equation}{section} \numberwithin{figure}{section}
\numberwithin{theorem}{section} \numberwithin{teorema}{section}

\def\be{\begin{equation}}
\def\ee{\end{equation}}
\def\bc{\begin{center}}
\def\ec{\end{center}}

\title{\bf A numerical investigation of the jamming transition in traffic flow on diluted planar networks}
\author{Adriano Barra\ $^1$,  Gabriele Achler\ $^2$}
\begin{document}
\date{}
\maketitle

\begin{center}
{\small \vskip-0.5cm


\footnote{e-mail:{\tt  adriano.barra@roma1.infn.it}} Dipartimento
di Fisica, Sapienza Universit\`a di Roma \vskip-0.5cm

\footnote{e-mail:{\tt  gabriele.achler@uniroma3.it}} Dipartimento
di Urbanistica, Universit\`a di Roma$3$ \vskip-0.5cm }
\end{center}

\vskip 0.5cm

{\small
\bf------------------------------------------------------------------------------------------------------------}

\vskip-1cm {\small

{\bf Abstract.} In order to develop a toy model for car's traffic
in cities, in this paper we analyze, by means of numerical
simulations, the transition among fluid regimes and a congested
jammed phase of the flow of {\em kinetically constrained} hard
spheres in planar random networks similar to urban roads .
\newline
In order to explore as timescales as possible, at a microscopic
level we implement an event driven dynamics as the infinite time
limit of a class of already existing model ({\em Follow the
Leader}) on an Erdos-Renyi two dimensional graph, the crossroads
being accounted by standard Kirchoff density conservations. We
define a dynamical order parameter as the ratio among the moving
spheres versus the total number and by varying two control
parameters (density of the spheres and coordination number of the
network) we study the phase transition.
\newline
At a mesoscopic level it respects an, again suitable adapted,
version of the Lighthill-Whitham model, which belongs to the
fluid-dynamical approach to the problem. \newline At a macroscopic
level the model seems to display a continuous transition from a
fluid phase to a jammed phase when varying the density of the
spheres (the amount of cars in a city-like scenario) and a
discontinuous jump when varying the connectivity of the underlying
network.
%
%
}

\vskip-0.6cm

{\small
\bf------------------------------------------------------------------------------------------------------------}

\section{Introduction}

In the past decades an always increasing interest has been payed
to the flow of {\em cars} in {\em urban} roads (see e.g.
\cite{intro1} or \cite{intro2} for a  beautiful modern review): a
primary challenge is the reduction of time delays and CO
emissions, in a nutshell, the congested traffic flow
\cite{inizio}.
\newline
Despite progresses developed by engineers and physicist (see i.e.
\cite{delbi,19}), essentially focused on large streets or small
tree-like graphs \cite{nega,naga}, very little is known concerning
the behavior of traffic on large two dimensional networks
\cite{intro2}.
\newline
In a completely different context, the last twenty years saw the
statistical mechanics of disordered and complex systems \cite{MPV}
experience an increasing development as well as its range of
applicability (see for instance \cite{guerrasg,alb2,amit}) and,
inspired by these successes, we want to investigate the nature of
the transition among a fluid state  and a jammed counterpart in
traffic flow on planar networks, investigating the presence (or
the lacking) of criticality \cite{24,barra6} within a out of
equilibrium statistical mechanics framework \cite{EeM,galina3}.
\newline
Furthermore, statistical mechanics of disordered systems recently
pointed out a deep connection among replica symmetry breaking
scenario \cite{barra10,MPV}, the paradigm of the transition among
fluid and glass and the $P \rightarrow NP$ transition in problem
solving of hard satisfiability problems \cite{mezard,pagnani}.
\newline
Interesting, if the mapping among {\em jamming transition} and
$P\rightarrow NP$ {\em completeness} would apply to traffic jams
too, it would vanish every attempt to an online control of car
flow by external massive macro-computing giving more firm ground
to  interacting local optimizers (as i.e. neural networks)
\cite{AI,peter,barraguerra}.
\newline
Deepening our knowledge concerning the jam transition in traffic
flow should be then of great importance, in our traffic
optimization planning, if, varying tunable parameters, glassy-like
criticality arises \cite{kurchan}.
\newline
From a practical viewpoint, as a rigorous formulation of out of
equilibrium statistical mechanics is far from being exhausted
\cite{gallavotti,kawasaki,EeM}, we do not have a paved
mathematical way to follow for checking i.e. the involved
time-scales \cite{Ton} or the reach of a stationary state
\cite{ciccottone} (which is a primary requirement for giving
meaning to the averages) and consequently there is the need of
fastest simulation algorithms  \cite{ciccottone2,barra3} to cover
as timescales as possible.
\newline
Even though we will move toward a molecular-dynamics-like approach
\cite{frenkel}, we stress that for similar reasons the biggest
amount of works on this subject uses in fact cellular automata
\cite{cellu,18,19,20} which are quite faster than the continuous
models \cite{10,11,81}.
\newline
Fast simulations are in fact  hard tasks, especially in models
with continuum potentials  as they need to be made discrete
generally by using Trotter expansions  of the Liouvillean
\cite{frenkel} describing the motion in the phase space,
forbidding a very long simulation time (i.e. by Liapounov
constrictions \cite{BeS}).
\newline
Avoiding these potentials, a very fast integration of the dynamics
is offered by the Verlet event driven dynamics of hard spheres
\cite{verlet}: these spheres are without a real potential; they
move on straight lines, up to a core distance at which they touch
one-other and they feel an infinite barrier of potential which
converts instantaneously the kinetic energy into potential energy.
\newline
In this way the motion against two successive collisions does not
require integration. It is in fact propagated from collision to
collision, the new positions and momenta are worked out  by
imposing conservation of particles (Kirchoff rule), energy and
momenta, (we will preserve just the Kirchoff rule in our
framework) and the motion is propagated again and so on
\cite{galina1} (we emphasize that this approach has been tackled
also to granular systems \cite{galina2}, which are glassy systems
 sharing several features with traffic flow
\cite{grano,granone}).
\newline
Of course there exist already several very sharp models for
traffic flow but we introduce our one because we are moving in an
opposite way for a different scope with respect the standard
approach: as we want a large amount of cars as well as long
simulation time for a thermodynamical approach, we allow ourself
to skip as details of the motion as possible, retaining just the
main features (as usually happens when looking at criticality in
statistical mechanics \cite{huang}).
\newline
Once defined the microscopic dynamics we then focus at first at
the mesoscopic level (order $10^2$ cars) to recover the
Lighthill-Whitham scenario \cite{50}, for showing consistence with
pre-existing works, then we focus on a macroscopic scale (up to
$10^4$ cars) to study its {\em thermodynamics}. We introduce the
ratio of the moving particle as a standard dynamical order
parameter \cite{intro2}, labeled by $\phi$ that we call {\em
fluidity} for the sake of clearness, and define it  as \be \phi =
1- N^{-1}\sum_i^N v_i, \ v_i \in [0,1] \ee (such that it is
trivially one in the jammed phase (where there is no longer any
motion) and decreases toward zero in the liquid phases) and study
its behavior: average, distribution and fluctuations.
\newline
The model seems to display a jammed phase where the fluidity is
strictly zero and its fluctuations are delta-like centered on the
average (the congested phase where all the spheres are caged among
their nearest neighbors) and a flowing phase in which the fluidity
seems to decrease continuously to zero (cages smoothly disappear)
by decreasing the density or discontinuously by increasing the
connectivity and its fluctuations appear Gaussians.
\newline
The whole suggesting the model undergoes a second order like
transition in the density and a first order like in the
connectivity.
\newline
For the sake of clearness we aim to label with $\alpha$ the
connectivity of the network,  with $\rho$ the density of the $N$
cars (even thought, from practical comparison  of different size
networks it will be sometimes easier to deal directly with the
un-normalized amount of car $N$) and with $v_i=v \ \forall i \in
(1,...,N)$ the velocity of the $i^{th}$ car.
\newline
It is worth noting that the standard technique of statistical
mechanics on diluted systems \cite{barra7,gt2} merges the two
parameters via the relation $ \alpha\tanh(\rho) = \rho'$, $\rho'$
being an equivalent density in a fully connected network, so
actually do not display clearly the transition split among the two
control parameters \cite{nota}.
\newline
Furthermore we want to stress that there exist already works on
dilute hard spheres in different contexts (as on the Bethe
lattice)  \cite{31,33} but, to our knowledge, not on networks with
topology close to urban one.

\section{Microscopic model}
In this section we point out the simulation scheme, which, for the
sake of simplicity we spit in two parts: the choice of the
underlying network (the topology) and the choice of the dynamics
on the network (the interactions).

\subsection{The Network}

At first we must introduce the graph.  In order to mimic a real
urban center we think at a graph whose links represent the roads
and vertices represent intersections and end points. For their
high connectivity, scale free \cite{caldarelli} and fully
connected \cite{hertz} networks are inappropriate to describe such
a graph and for the extremum order and homogeneity they present,
also Voronoi tessellation \cite{voronoi} and regular grids are
avoided \cite{grid}.
\newline
Real data show a linear dependence among the number of roads
versus the number of intersections \cite{latora}, whose ratio must
be obviously between one (tree like structures with no loops) and
two ($2D$ regular lattice) with an empirical slope close to $3/2$
\cite{flammini}. Even though clever growth algorithms for these
networks recently developed \cite{flammini}, for computational
simplicity (as we will have to average over several configurations
and we want the fastest procedure) we choose the planar
Erdos-Renyi graph \cite{caldarelli} above the giant component
threshold, which is close to the requested class of random graphs
\cite{erdos2D} and is of immediate realization on a computer as no
growing is concerned.
\newline
All the links represent streets built by two lanes so to have both
an incoming and an outgoing flux from each node.
\newline
On this network we can vary its averaged connectivity -the
coordination number- (denoted by $\alpha$) so to explore from the
region of extreme dilution near the percolation threshold up to a
fully ordered grid.
\newline
Of course to check convergence to the infinite volume limit we
will test our simulations varying also the size of the grid.

\begin{figure}
\begin{center}
\includegraphics[angle=-90, width=5.5cm
]{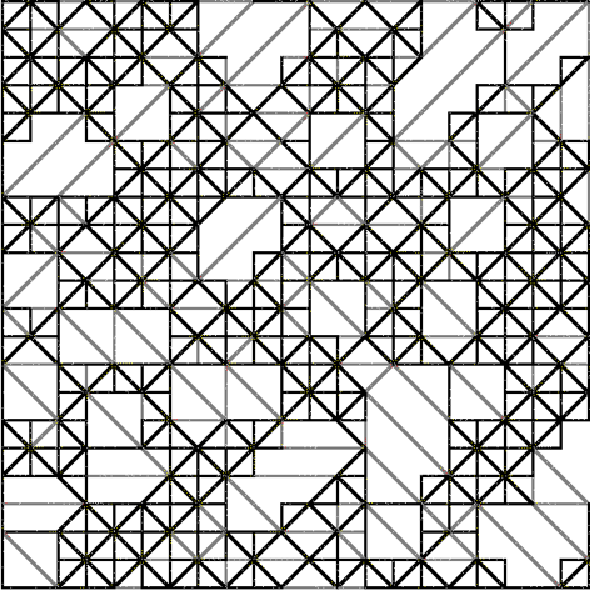} \ \ \includegraphics[angle=-90, width=5.5cm
]{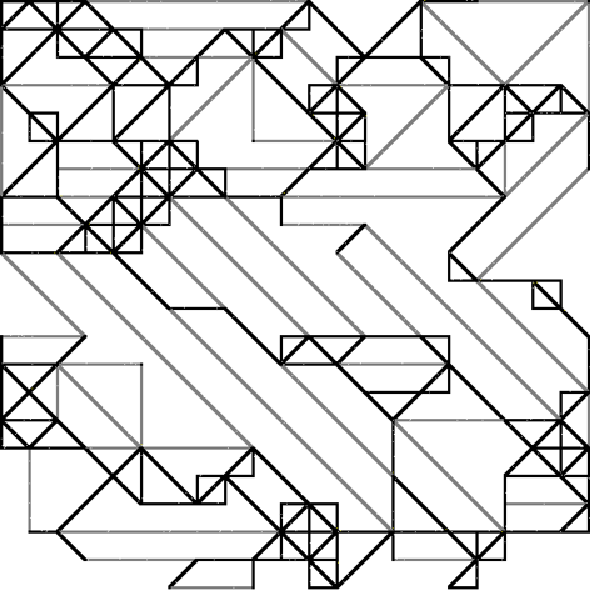}
\end{center}
\caption{\itshape Examples of generated networks. At left an high
connectivity topology, at right a low one. We start from a fully
ordered grid in which each node has exactly $8$ links then we
dilute the links with a Poisson distribution with mean $4$, such
that the main nodes are ordinary crossroads. If a node has two,
one or zero links is removed too.}\label{rete}
\end{figure}

\subsection{The dynamics}

On this network we let live $N$ {\itshape cars} for which we want
to menage two limits at the same time (which conflict in terms of
CPU time): the {\itshape large $N$ limit} to have enough data for
the averages and the {\itshape infinite time behavior} so to
approach to a steady state: the need of the fastest plausible
algorithm follows straightforwardly.
\newline
One of the most important class of models which aim to mimic car
dynamics is the so called {\em follow the leader}
\cite{boh1,fdl,nega}, where the dynamics of the $i^{th}$ car is
assumed to respect the following differential equation (or some
suitable variants \cite{intro2}), where $(i+1)^{th}$ labels the
car in front of the $i^{th}$, following the direction of motion:
\be\label{ftl} \frac{d^2 x_i}{dt^2}\sim (v_i - v_{i+1}). \ee In a
nutshell the car $i$ accelerates if the car in front of is
accelerating too (this happens both for positive and negative
accelerations).
\newline
As a solution for the long time steady state behavior of the
{\itshape follow the leader} model is given by the same constant
velocity for all the cars (as it appears clearly by solving
eq.(\ref{ftl}) in the large time limit by imposing $d^2 x_i / dt^2
= 0$) we choose for our microscopic dynamics an event driven
hard-sphere-like dynamics \cite{galina1}, which is known to be one
of the fastest achievable dynamics in terms of CPU time
\cite{barra3,elber} and is in agreement with these continuous
potential in the regime where we are interested in ($t \rightarrow
\infty$).
\newline
On every lane overtaking is forbidden (FIFO principle \cite{fdl})
and every car has an energy $E_i= v^2_i/2$ during the free motion,
whose dynamics is propagated between collisions among two cars or
one car and a crossroad.
\newline
There the collision rules are worked out with a remarkable
difference with respect to canonical physics: nor detailed balance
neither the third law of dynamics do hold. When a collision
happens there is no conservation of momenta and the incoming car
loses all the energy then starts again following the collided car
(crossroads apart which confer a certain degree of randomness),
for this reason we call our hard spheres {\em kinetically
constrained} \cite{KCM1,KCM2}.
 However as for instance
elegantly explained in \cite{22,galina3} this is not a too serious
limitation when investigating out of equilibrium steady state (the
same is not true of course in equilibrium \cite{huang}).
\newline
The potential felt by the cars is the classical hard core
potential \cite{galina1,barra3} $V$, which, by using $|x_{ij}|$ to
evaluate the distance between the two cars situated respectively
at $x_i$ and $x_j$, can be written as
\begin{eqnarray}\label{potenziale}
V (|x_{ij}|) \equiv \left\{
\begin{array}{rl}
V_0 = 0 \ \ \ \ \rm{if} \ \ \ \ |x_{ij}| \geq d \\
V_1 = \infty \ \ \ \ \rm{if}  \ \ \ \  |x_{ij}| < d
\end{array}
\right\}
\end{eqnarray}
where $d$ can be thought of as a {security distance}, the minimal
distance allowed among two consecutive cars (we stress that
varying $d$ changes also the maximum number of admitted cars
inside a network $N_{max}$ -which sets $\rho=1$- and consequently
the critical density for the transition, so we fix $d=1$ once for
all).
\newline
So far we defined the network and the dynamics along a link
(street); we must further specify what happens when more links
merge in a node (crossroad): several decision rules can be
implemented; in this preliminary work we impose a simple random
walk at the nodes: if a car is at the end of a link and has $n$
possible directions to take it chooses one of them with
probability $n^{-1}$.
\newline
Of course the total number of machines is conserved along the
dynamics and we impose Kirchoff rules \cite{intro2} for the flow
at the nodes.

\section{The adapted Lighthill-Whitham theory}

As we want to move from a microscopic prescription toward a
macroscopic description, there should be a  {\itshape mesoscopic}
lengthscale at which well known models, the most famous being the
Lighthill-Whitham model (LW) \cite{50}, should be recovered. With
{\itshape mesoscopic} we mean we are dealing with an ensemble of
cars such that their concentration as a function of the space
(labeled by $x$) and time (labeled by $t$) is meaningful whilst
the average overall the concentrations can still be thought of in
terms of a probabilistic description \cite{SGR}.
\newline
From a practical viewpoint let us now concentrate just on a big
street with an amount $M$ of cars inside and analyze the flow at
this level.
\newline
Assuming that the cars move from smaller to larger values of $x$,
we define the concentration $C$ of an element of the traffic on
the street as the amount of cars moving within the space delimited
by two generic points $C \supset x \in [x_a,x_b],\ x_a < x_b$. It
follows that $C(x,t)=M^{-1}\sum_i^M
(\theta(x_i(t)-x_a)-\theta(x_i-x_b))$ and let us also  write the
velocity of the generic $i^{th}$ particle as $v_i(t)=1 -
\delta((x_i(t)+d)-x_{i+1}(t))$, such that if the car is moving,
its velocity is one, otherwise is zero.
\newline
The traffic flow  is defined as $J(x,t)= v_g(x,t) C(x,t)$ where
$v_g(x,t)$ is the group velocity and $C(x,t)$ the concentration.
Assuming conservation on the total amount of cars, the following
continuity equation holds \cite{50} \be\label{cazzo}
\frac{\partial C(x,t)}{\partial t} + \frac{\partial
J(x,t)}{\partial x} = \sum_i^N \rho^i_{in}(x,t) -\sum_{j}^N
\rho^j_{out}(x,t) \ee where $\rho^i_{in}(x,t) =
\theta(x_a-x_i(t))-\theta(x_a-x_b-x_i(t))$ and $\rho^i_{out}(x,t)=
1 - \rho^i_{in}(x,t)$ take into account car entries and exits on
the street.
\newline
The group velocity relates to the local velocity of the particles
via \cite{intro2} $v_g= \frac{1}{M}\sum_i^Mv_i + \frac{1}{M} C
\partial_C \sum_i^M v_i \leq 1$ which in our case reads off as \be
v_g = \langle v_i \rangle - \big\langle \frac{1}{N}\sum_i^N v_i
\big\rangle = 1 - \phi,\ee where the brackets average out the
Dirac deltas on some measure (note that $v_g \leq \langle v_i
\rangle$).
\newline
The scenario is as follows: as far as the system is completely
un-frustrated (low density liquid) the group velocity corresponds
to the single velocity; increasing density of cars (or decreasing
the connectivity of the network) frustration arises and lowers the
group velocity up to a threshold where the jamming transition
 starts and $v_g$ goes to zero such that the only solution to eq.
(\ref{cazzo})  is $C(x,t) = C \in [0,1]$.
\newline
For the sake of simplicity let us consider a very long piece of a
street without entries or exits such that the LW model in our case
obeys the following partial differential equation \be
\frac{\partial C(x,t)}{\partial t} + (1- \phi)\frac{\partial
C(x,t)}{\partial x} = 0, \ee whose solutions are Galilean
characteristics with slope $1-\phi$ and can be expressed by
generic functions $f(x \pm (1-\phi) t)$ (cinematic waves
\cite{51}). If we now focus on two adjacent elements,
$C_1(x,t),C_2(x,t)$ and suppose that $C_1(x,t)$ is in front of
$C_2(x,t)$ in the direction of motion, we see that as far as their
corresponding fluidities are zero they both follow straight lines
(on the $x,t$ plane) with the same slope. If we now suppose that
$C_1$ changes its status (for example by internal rearrangements,
that we impose in the simulation by freezing randomly spheres
belonging to the first package) and becomes frustrated $\phi_1>0$,
its group velocity decreases and at the time $\Delta X /
(v_{g1}(\phi)-v_{g2})$ the two characteristics meet causing a
discontinuity in the concentration functions, which acts as the
starter of the transition, as (due to the structure of the
solution) it propagates both forward and backward on the street.
\begin{figure}
\begin{center}
\includegraphics[width=5cm
]{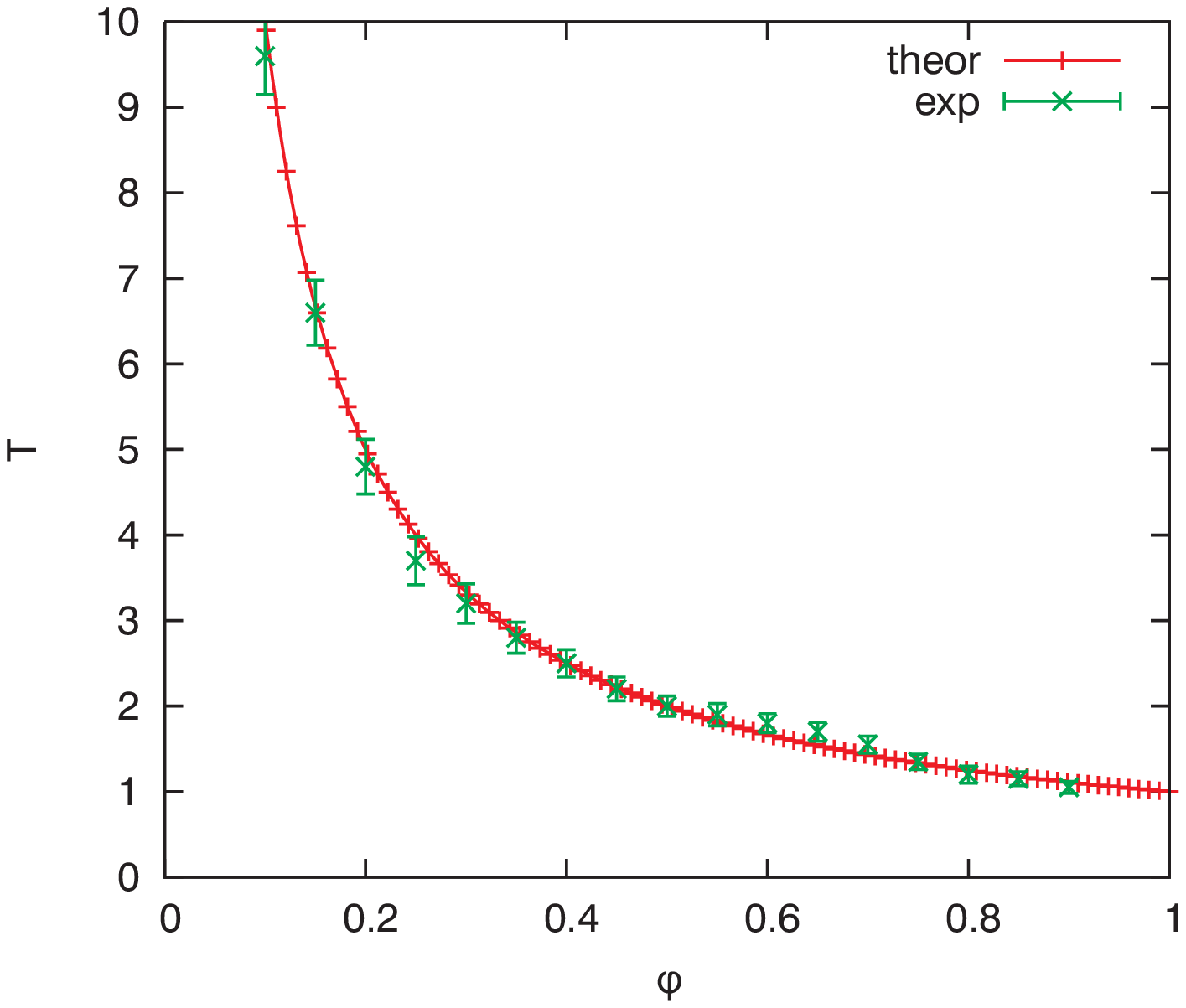} \ \ \includegraphics[width=6cm,
]{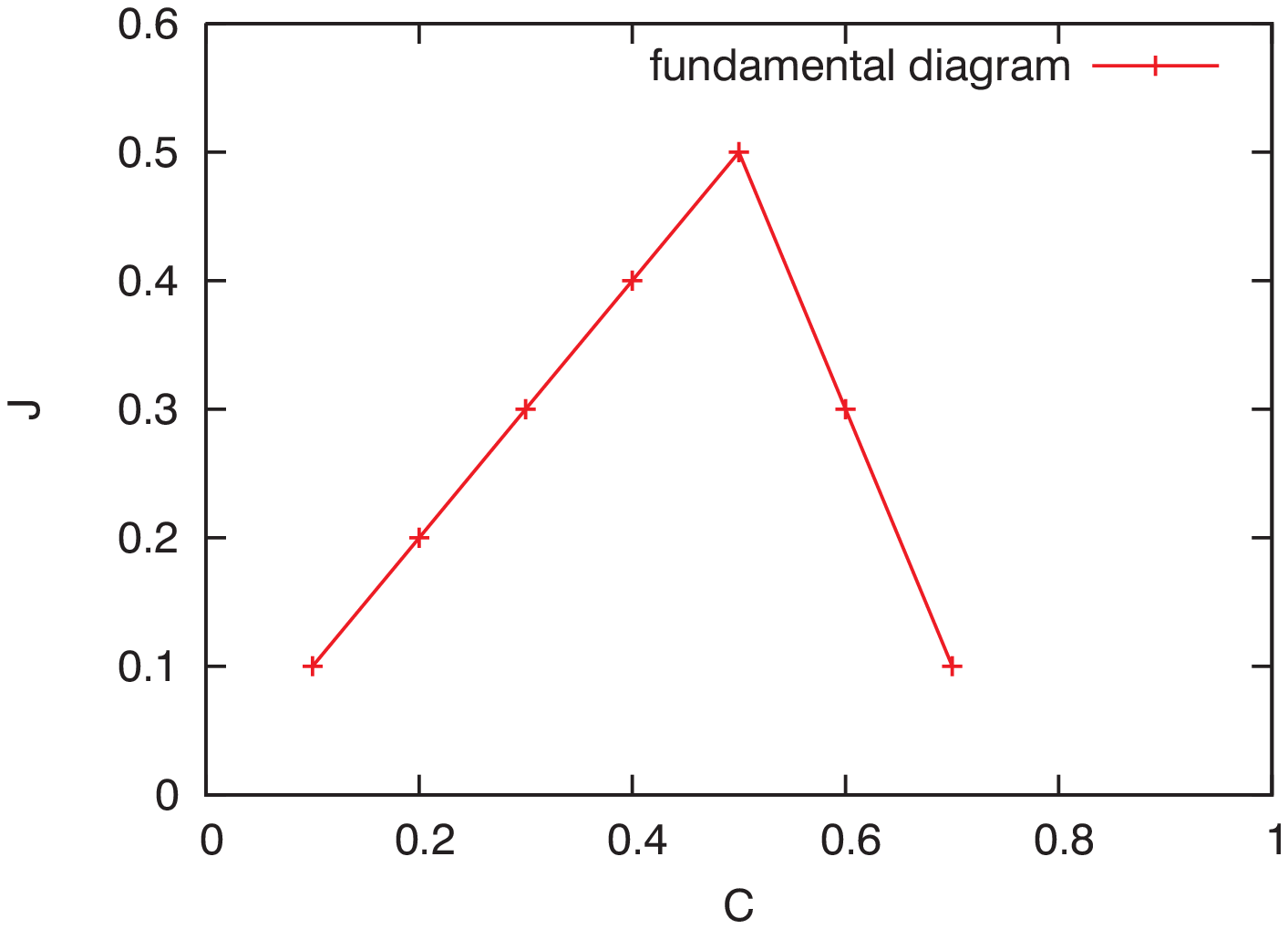}
\end{center}
\caption{\itshape Left: Collision time versus the fluidity of the
first concentration $C_1$ in our LW model, {\em theor} stands for
analytical prediction, {\em exp} for the simulation results.
Right: Fundamental diagram for our LW model.
}\label{meso}
\end{figure}

\section{Macroscopic behavior}

\begin{figure}
\begin{center}
\includegraphics[width=4cm
]{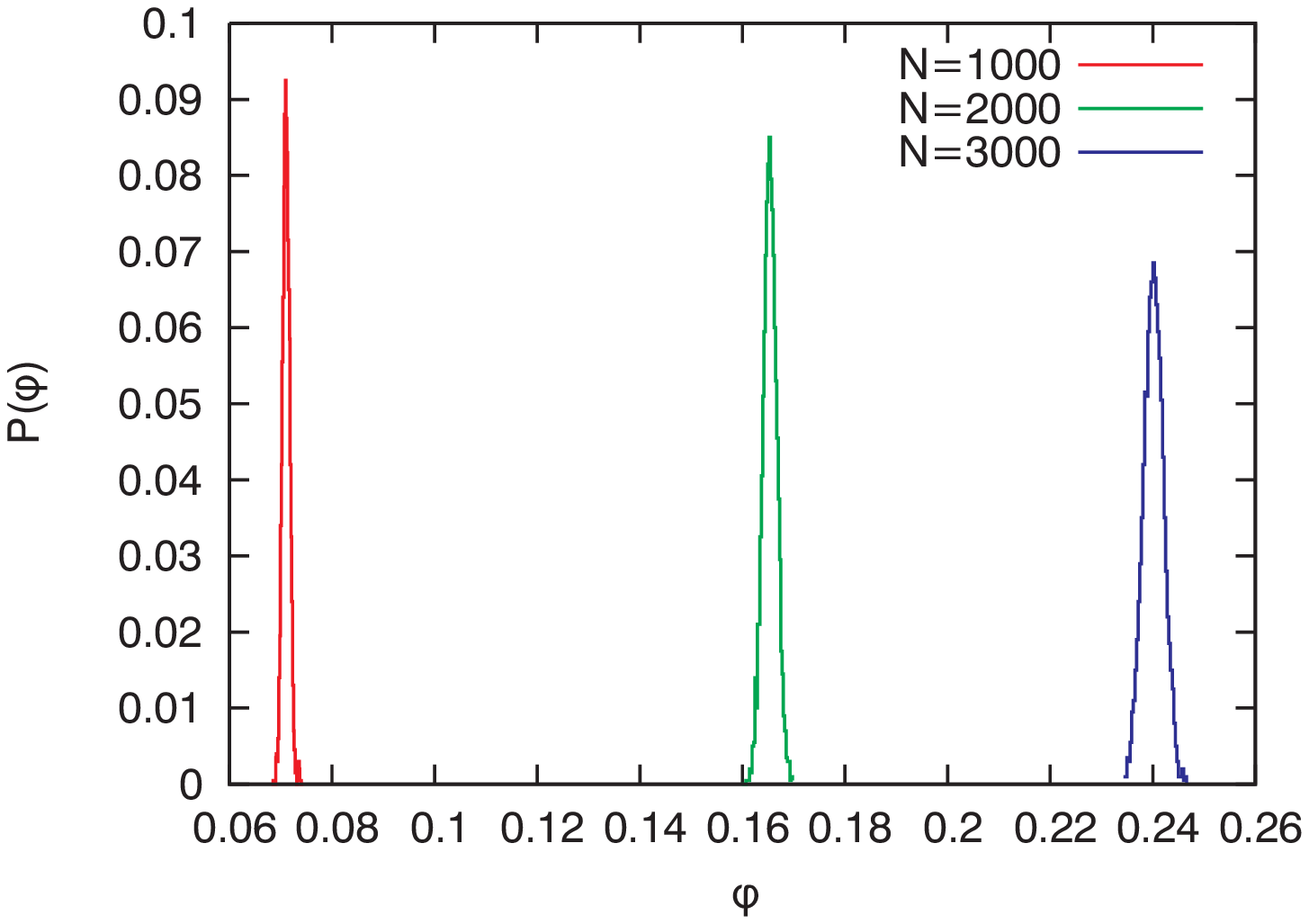}\includegraphics[width=4cm
]{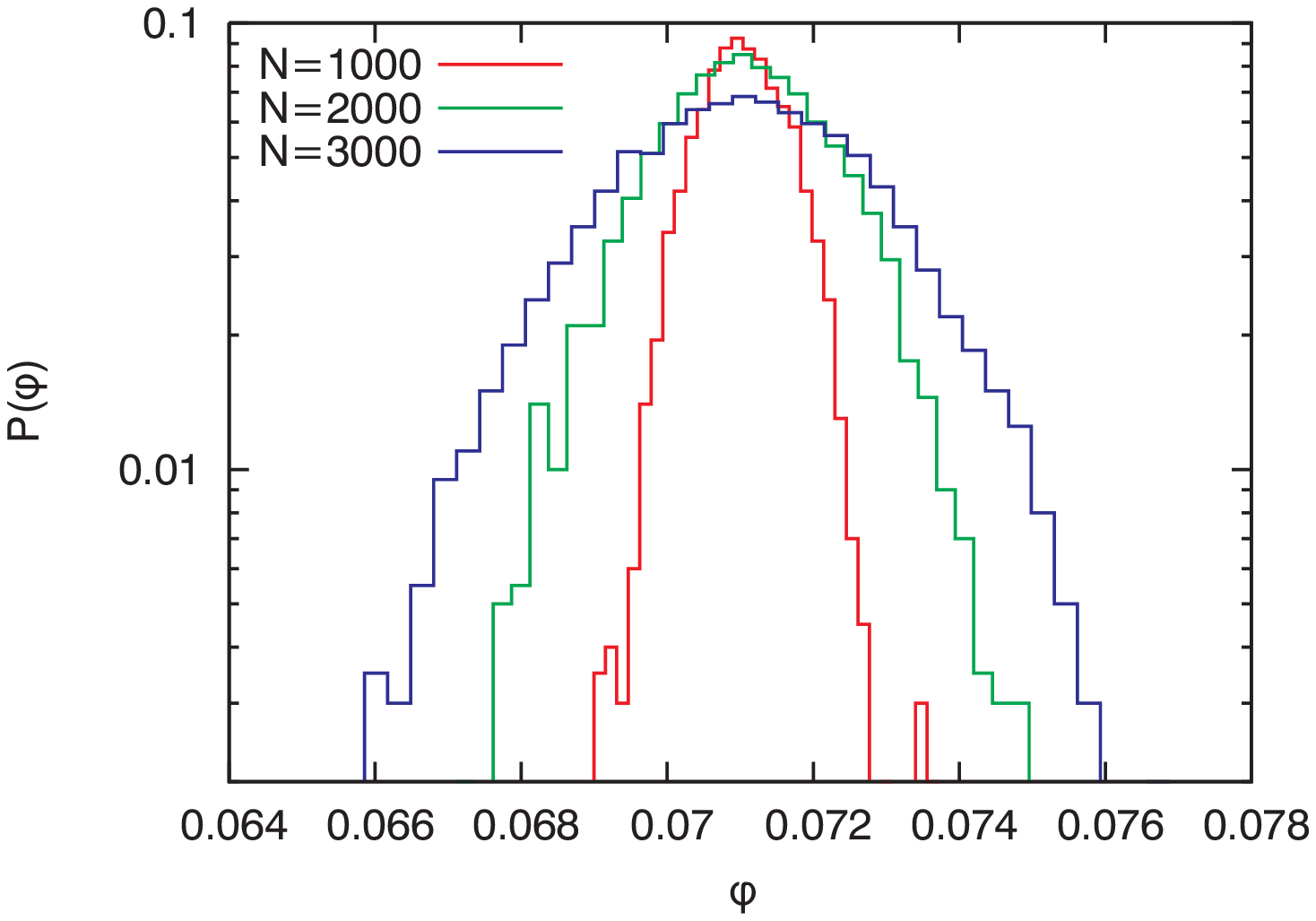}\includegraphics[width=4cm
]{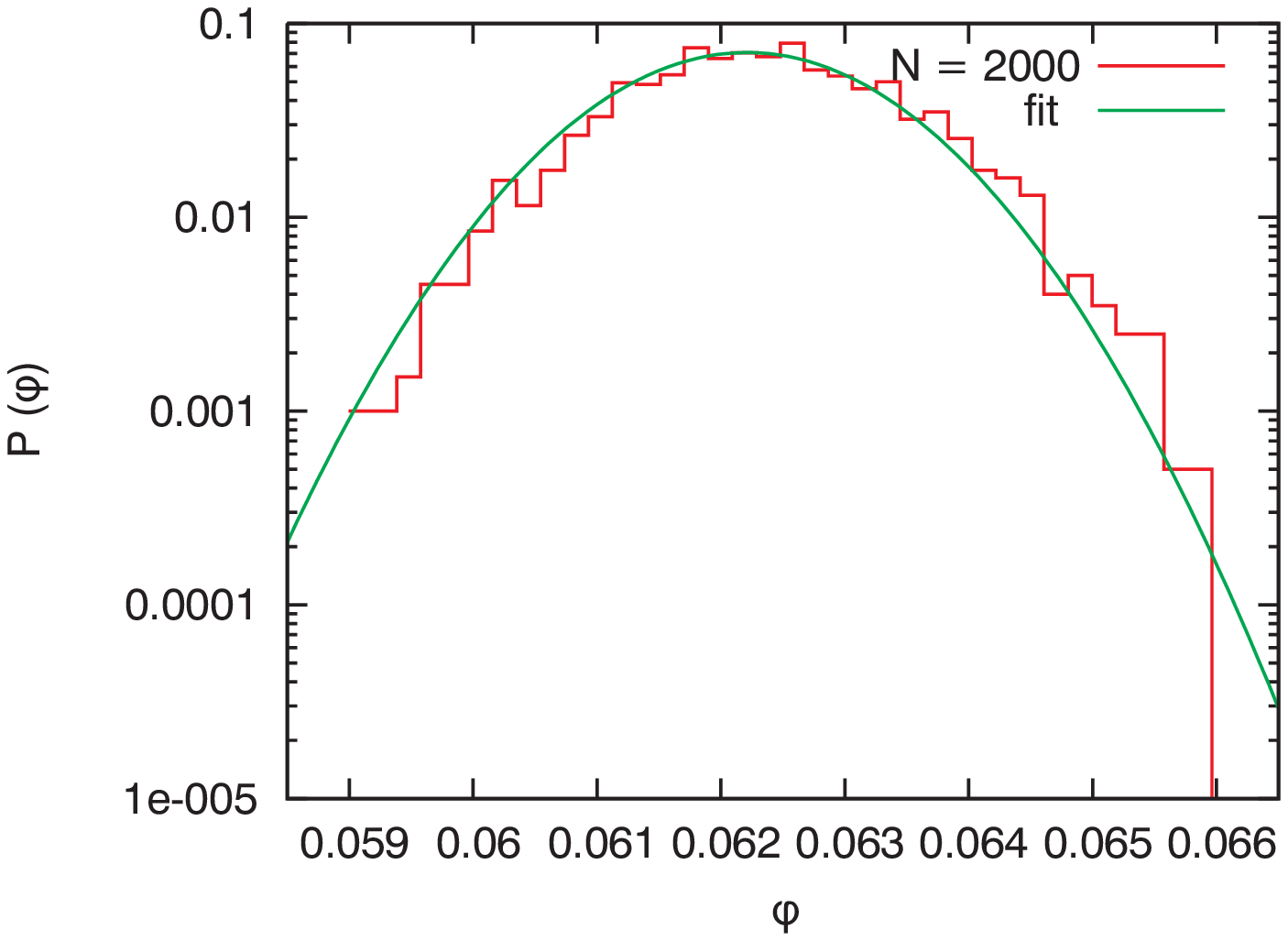}
\end{center}
\caption{\itshape Fluidity fluctuations in the steady state for
the largest size city at a connectivity $\alpha=0.6$: We show the
distributions of $3$ different densities ($N=1\cdot10^3,
2\cdot10^3,3\cdot10^3$) on a lin-lin scale (left), the spread of
their variances, centering the distributions on the smallest
average for easier visualization (center) and a Gaussian fit
$(\chi^2 \sim 0.93)$ on a log-scale obtained by sampling over
$2000$ cars.}\label{fuck}
\end{figure}

\begin{figure}
\begin{center}
\includegraphics[width=6.3cm
]{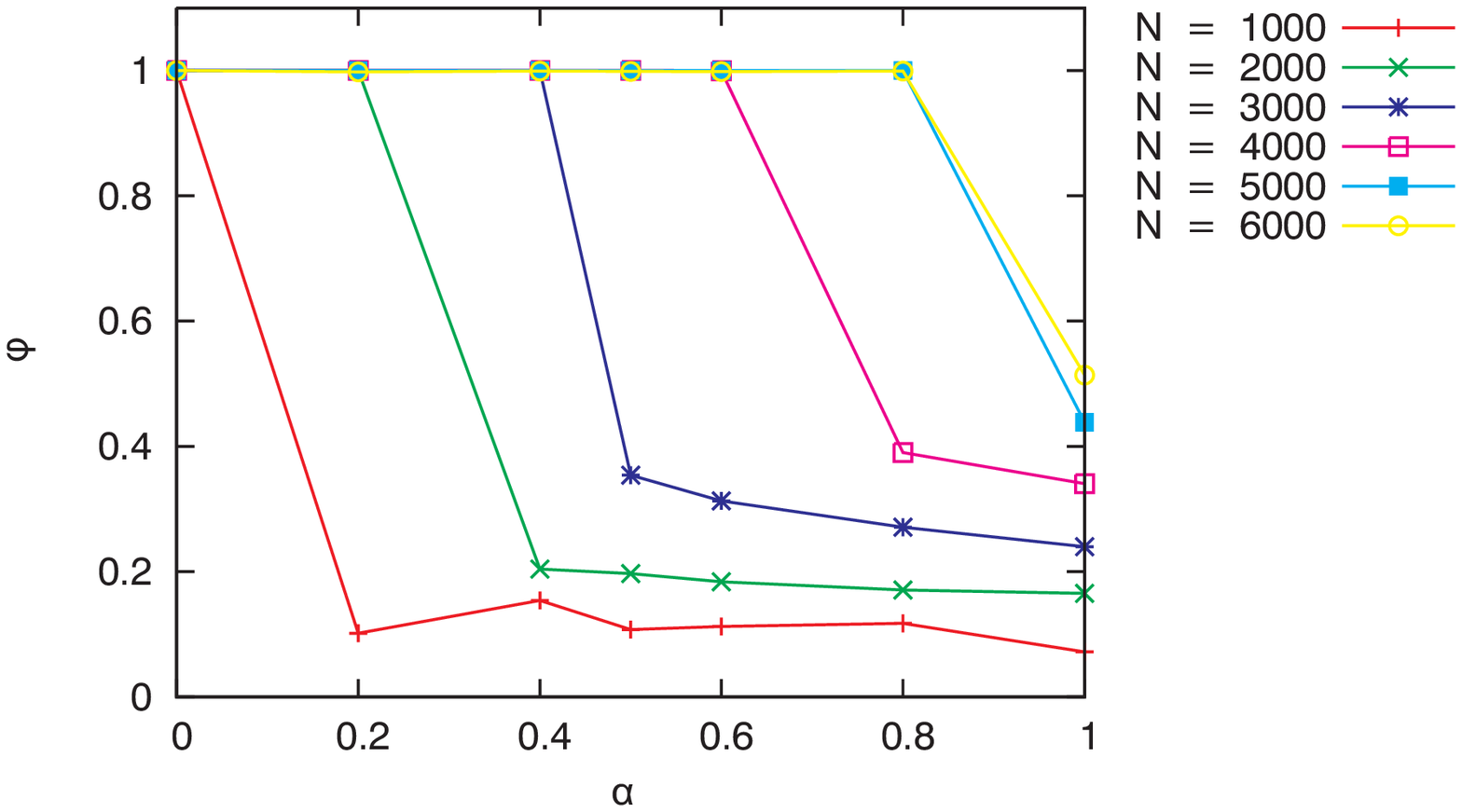}
\includegraphics[width=5cm
]{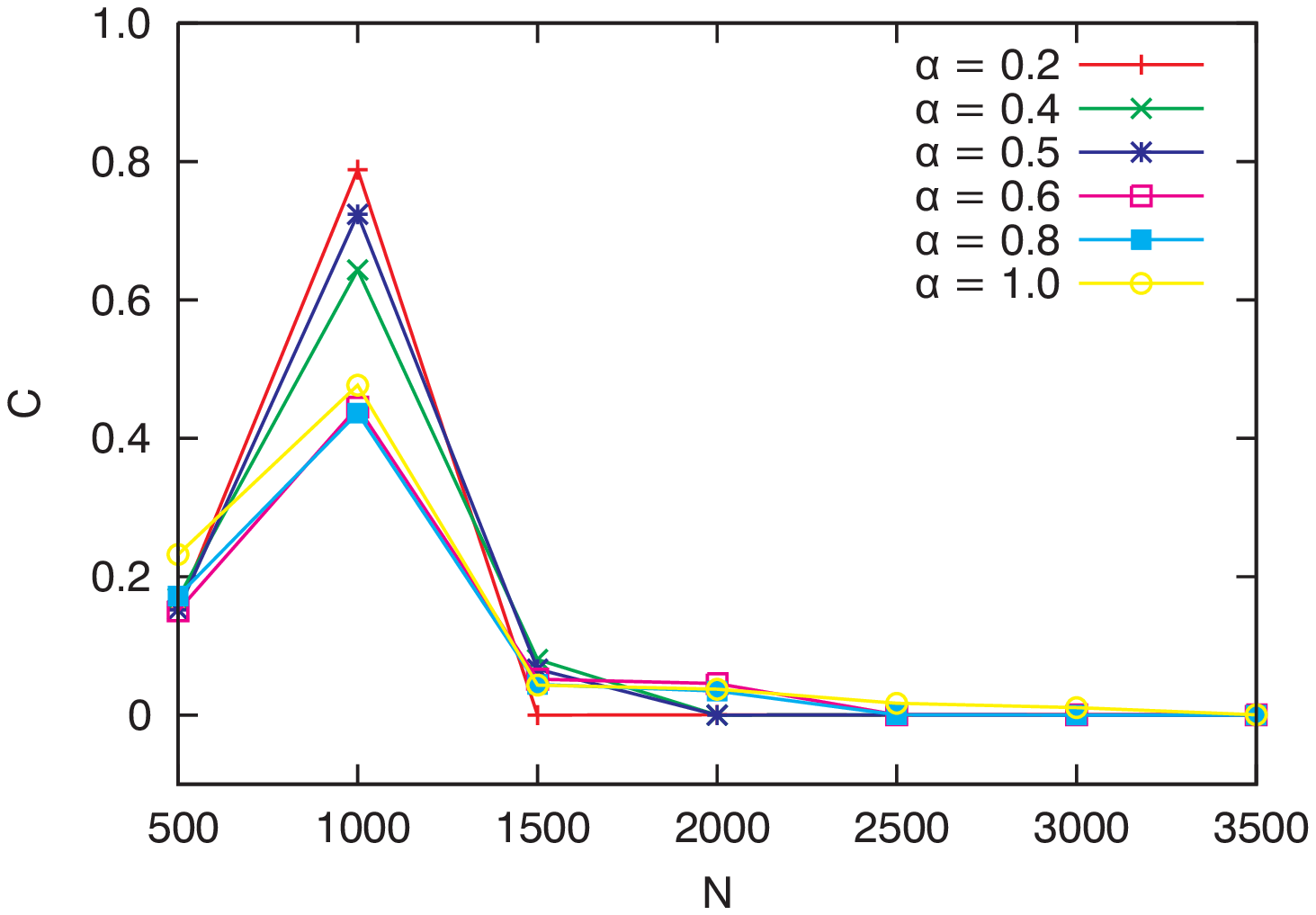}
\end{center}
\caption{\itshape Left: Medium city. Behavior of the fluidity
versus the connectivity for five different densities, shown in
terms of the total amount of cars $N$. Right: Smallest city.
self-averaging $C$ of the order parameter: fluctuations versus the
density (expressed in terms of the total amount of cars), for six
different connectivity values. The onset of the jam transition is
shown by the peak, which, increases proportionally to
$\alpha^{-1}$.}\label{conne}
\end{figure}

Now we turn to the analysis on large scale networks. We consider
three sizes of (squared) networks, which, at the highest
connectivity, are completely filled by $N_{max}=2\times10^3$,
$5\times10^3$ and $1\times10^4$ spheres, which set $\rho=1$ in the
different cities.
\newline
The connectivity ranges from $0$ to $1$ and is changed from the
percolation threshold (which is defined by $\alpha=0$) to the
fully two dimensional network (which is defined by $\alpha=1$) in
eight steps and the density (again defined in $[0,1]$) is
investigated at five different values too.
\newline
The analysis works as follows: for every investigated size, degree
of connectivity and density, we average over twenty different
runs; for every run we start off spheres randomly over the network
(avoiding pathological overlaps)  and study the mean of the
fluidity, its variance and its distribution by starting to collect
data when the mean does not vary sensibly (less than $5\%$ for an
amount of time $O(10^5)$ collisions), that we claim to be a
stationary state.
\newline
\newline
At first, even though we do not provide a plot (as it would have
just three points), we stress that we obtained good agreement
among the three different city sizes investigating
 the monotonicity of the convergence of the averages (of fluidity and its variances), which is
the basis of the thermodynamic limit (fundamental property for a
well behaved model \cite{limterm}).
\newline
For all the not-jammed stationary states (at fixed $\alpha,\rho)$
we analyzed the distribution of the fluidity sampling over the
whole simulations performed: they turn out to be close to
Gaussians distributions (which we use as a test fit, see figure
(\ref{fuck})), the variance being almost independent by $\alpha$
and increasing with the density of the spheres up to the
transition point (at given $\alpha$), immediately later they are
delta-like on the averaged of the fluidity ($\phi = 1$).
\begin{figure}
\begin{center}
\includegraphics[width=5cm
]{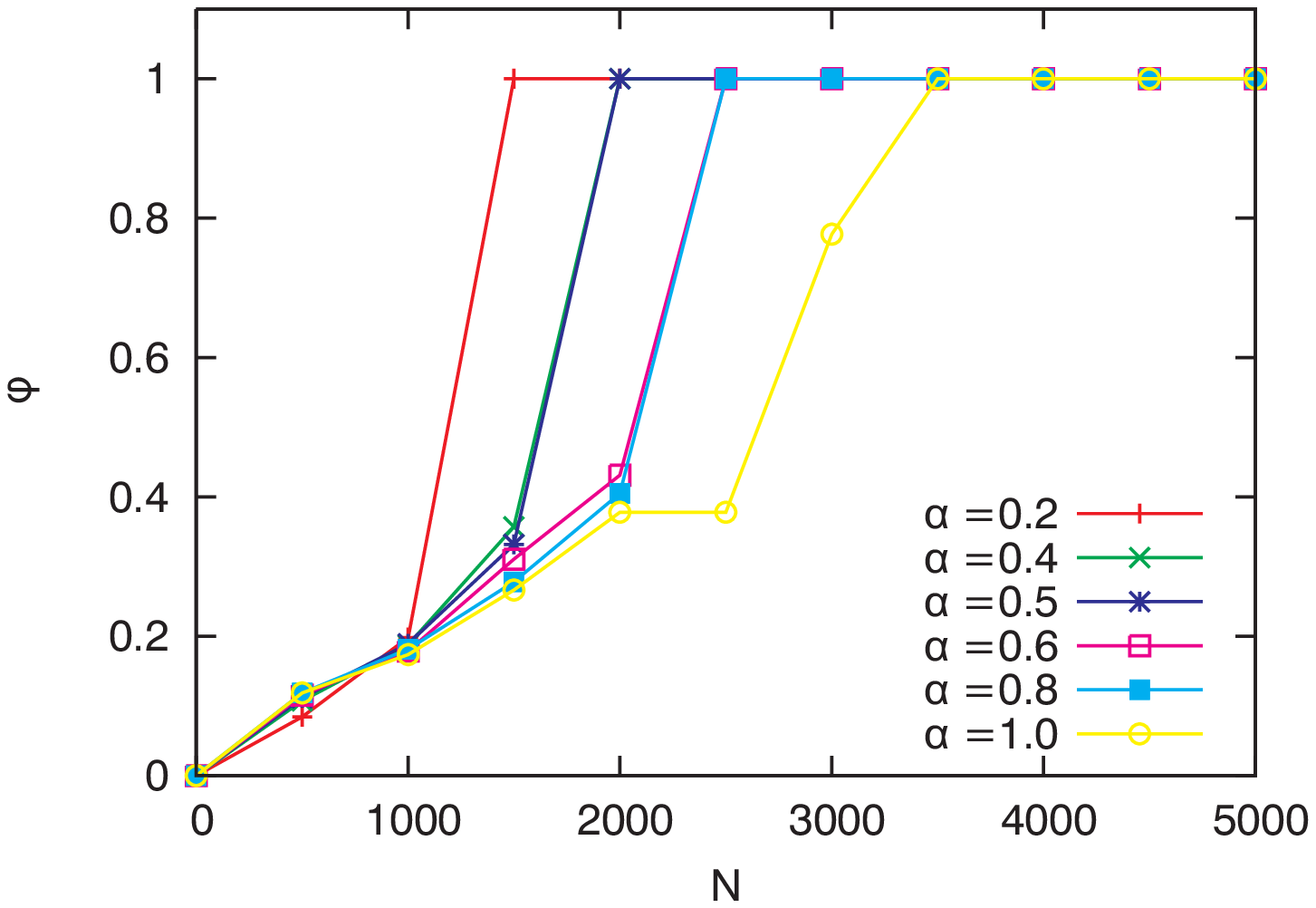}
\includegraphics[width=7cm
]{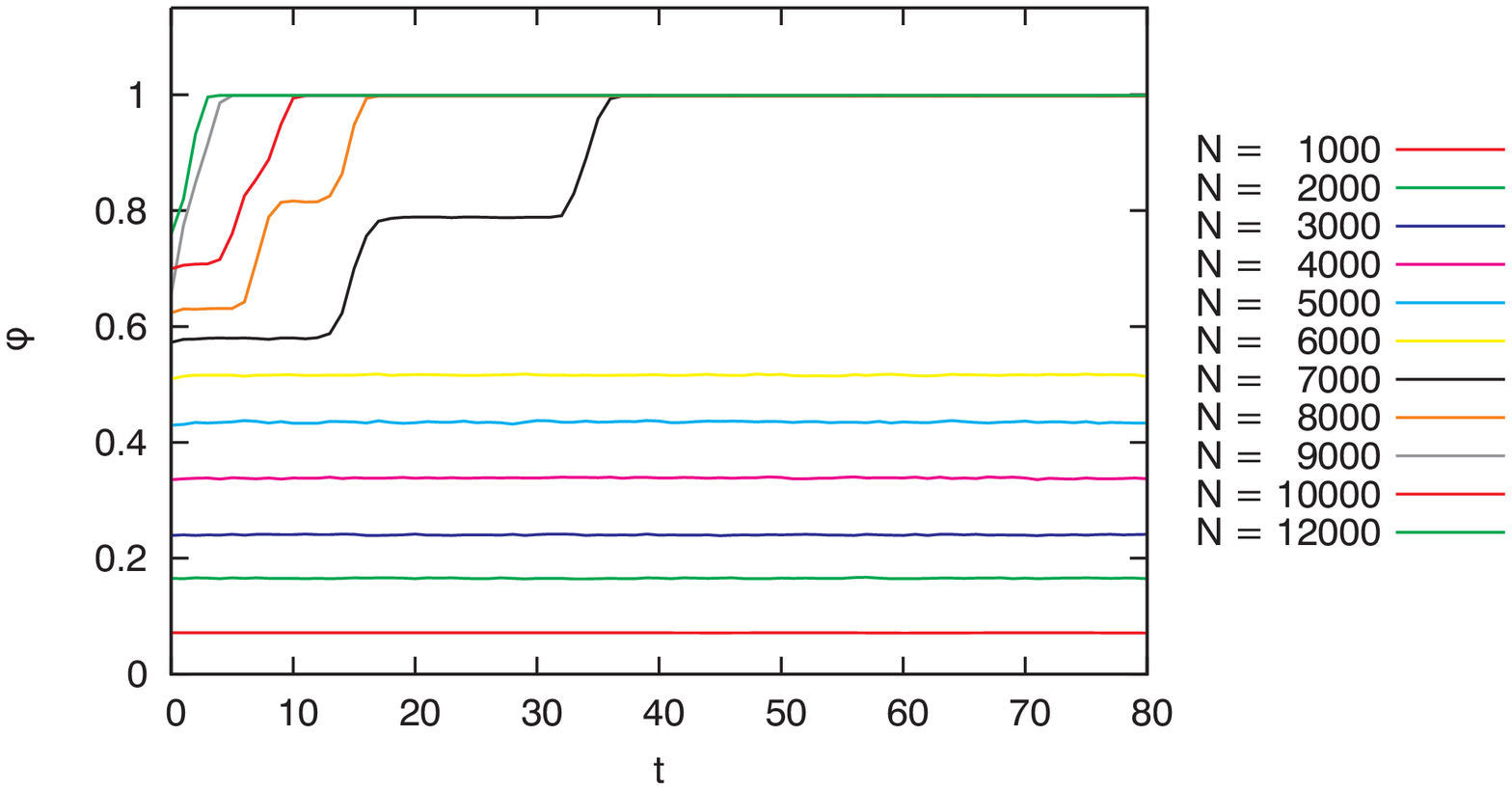}
\end{center}
\caption{\itshape Left: Behavior of the fluidity versus the total
number of cars at six different values of the connectivity. In the
low density regime the dependence of $\phi$ versus $N$ is linear
and universal. Right: Example of fluidity versus time on a single
network (time is measured by the number of collisions (modulus
$10^4$) showed from the larger city: there exist only two
"macroscopic" long term behavior: a jammed phase, that for the
investigated values appears for $N \geq 7\times10^3$ and a liquid
phase, which however shows a continuous variation in the fluidity
versus the density of cars (for $N < 7\times 10^3$).}\label{timeo}
\end{figure}
The average fluidity versus the density (that we plot directly by
using the amount of cars) is shown in figure (\ref{timeo}): For
low density regimes, flow behavior appears independent by the
connectivity of the network (and the scaling is always the same
$\phi \propto \rho$), while, for higher values of the density a
second regime is approached which is highly sensible by the
connectivity, and in which, continuity of the order parameter with
respect to the density is still observed.
\newline
Finally, for highest level of density a discontinuous jump to the
jammed phase is observed, for all the values of the connectivity
(see figure (\ref{conne}) where we report results for the medium
city size).
%
%

In figure (\ref{conne}) furthermore we show the fluctuations of
the order parameter for the smallest city: it is worth noting that
a phase transition (marked by a sharp peak inversely proportional
to the connectivity) seems to appear at a critical value of
density.
Analyzing again the medium size city, we show in figure
(\ref{conne}) the behavior of the order parameter versus the
connectivity (for several values of density) and there is no
presence of a continuous behavior: at a critical value (depending
on the density) a jump to a jammed phase is observed.
\begin{figure}
\begin{center}
\includegraphics[width=9cm
]{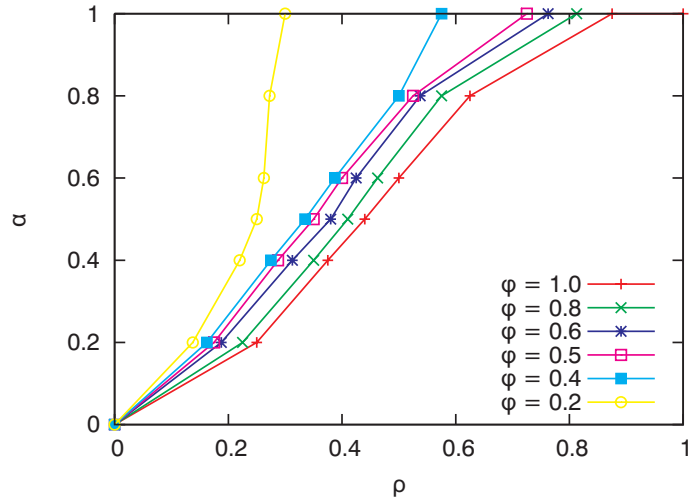}
\end{center}
\caption{\itshape Phase diagram: The line $\phi=1$ is the
transition line. At right there is the jammed phase at left the
fluid phase. Further, the latter is shown split in six zones
accordingly with different values of the fluidity.}\label{densa}
\end{figure}

\section{Conclusions}

In this paper we implemented a numerical algorithm which mimics
the flow of cars in urban cities: cars are described as {\em
kinetically constrained} hard spheres and urban topology is chosen
as a planar Erdos Renyi graph. {Kinetically constrained} because
the one way roads break detailed balance and the collisions among
two cars do not respect the third law of thermodynamics. Even
though mathematically hard to be analyzed, this model can still be
investigated by numerical simulations. Hard spheres because, as in
principle we do not know how many timescales are involved in the
genesis of the congested phase, and being interest in the long
time behavior, we have chosen one of the fastest possible
algorithms for the dynamics: the event driven motion, which
requires hard spheres. We investigated the response of a dynamical
order parameter, the fluidity, defined as the ratio among the
moving cars on the whole ensemble, by tuning two control
parameters: the density of the cars and the connectivity of the
network.
\newline
From this numerical investigation we found a continuous transition
from a congested phase to a fluid phase by varying the density of
the cars (at fixed connectivity) such that the fluidity lowers
smoothly from $1$ to smaller values (up to zero where there are no
longer cars) and a discontinuous jump of this order parameter when
varying the connectivity of the network at fixed amount of cars.
\newline
Furthermore the timescales involved seem to be several (the
longest of which seems to diverge at the transition to a jammed
phase) rising the question on what kind of traffic optimizer
should be developed in order to minimize traffic.
\newline
On these first heuristic considerations we believe that
interacting local optimizers (as a grid of  interacting
cross-lights able to detect flow \cite{fdl,finale}) would work
better than a global ground state searcher and are more stable
with respect to perturbations as new added (or removed) streets.
\newline
Future works concerning the kind of transition will be due to
investigate the relaxation to equilibrium after the stimuli by
introducing a car (or a few) or by introducing a new link, so to
check the presence of aging in the network. Furthermore traffic
 optimization by properly interacting traffic lights will be considered as
well.

\section*{Acknowledgements}

The authors are grateful to Francesco Guerra, Paolo Avarello,
Viola Folli, Roberto D'Autilia  and Elena Agliari for useful
discussions. AB work is partially supported by the SmartLife
Project (Ministry Decree $13/03/2007$ n.$368$) and partially by
the
 CULTAPTATION Project (European Commission
contract FP6 - 2004-NEST-PATH-043434).



\begin{thebibliography}{0}
\bibitem{nota} Strictly speaking the investigated systems we meant
are spins on lattice, whose order parameter is the temperature,
which is well known, for model with discontinuous potentials,
substituted by the density.
\end{thebibliography}
\end{document}